# Visually Comparing Graph Vertex Ordering Algorithms through Geometrical and Topological Approaches

**Karelia Salinas** [ *Institute of Mathematical and Computer Sciences, University of São Paulo* | *karelia@usp.br* ]
**Victor Barella** [ *Institute of Mathematical and Computer Sciences, University of São Paulo* | *victorhb@icmc.usp.br* ]
**Thales Vieira** [ *Institute of Computing, Federal University of Alagoas* | *thales@ic.ufal.br* ]
**Luis Gustavo Nonato** [ *Institute of Mathematical and Computer Sciences, University of São Paulo* | *gnonato@icmc.usp.br* ]

*Institute of Mathematical and Computer Sciences, University of São Paulo, Av. Trabalhador São-carlense, 400, Centro, São Carlos, SP, 13566-590, Brazil.*

**Abstract.** Graph vertex ordering is a resource widely employed in spatial data analysis, particularly in the urban analytics context, where street graphs are frequently used as spatial discretization for modeling and simulation. Vertex ordering is also important for visualization purposes, as many methods require the vertices to be arranged and displayed in a well-defined order to enable the visual identification of non-trivial patterns. The primary goal of vertex ordering methods is to find an ordering that preserves neighborhood relations. However, the structural complexity of graphs employed in real-world applications leads to unavoidable distortions in the ordering process. Therefore, comparing different vertex ordering methods is fundamental to enable effective analysis and selection of the most appropriate method in each application. Although several metrics have been proposed to assess spatial vertex ordering, they typically focus on measuring the quality of the ordering globally. Global ordering assessment does not enable the analysis and identification of locations where distortions are more pronounced, hampering the analytical process. Visual evaluation of the vertex ordering mechanisms is particularly valuable in this context, as it allows analysts to distinguish between methods based on their performance within a single visualization, assess distortions, identify regions with anomalous behavior, and, in urban contexts, explain spatial inconsistencies in the ordering. This work introduces a visualization-assisted tool to assess vertex ordering techniques, having urban analytics as the application focus. Specifically, we evaluate geometric and topological vertex ordering approaches using urban street graphs as the basis for comparisons. The visual tool builds upon existing and newly proposed metrics, which are validated through experiments on urban data from multiple cities, demonstrating that the proposed methodology is effective in assisting users in selecting a suitable vertex ordering technique, fine-tuning hyperparameters, and identifying regions with high ordering distortions.

## 1 Introduction

Vertex ordering is a fundamental problem in graph analysis with significant implications for applications such as visualization [Behrisch *et al.*, 2016], word cloud generation [Paulovich *et al.*, 2012], finite element methods [Kaveh and Bondarabady, 2002], cache optimization [Wei *et al.*, 2016], and mesh generation [Berger *et al.*, 2010]. In spatial graphs, where nodes represent geolocated entities, ordering plays an important role in preserving spatial relationships and ensuring meaningful data representations in analytical processes. Vertex ordering enables effective visual exploration and pattern identification, facilitating tasks such as spatio-temporal data visualization [Franke *et al.*, 2021; Buchmüller *et al.*, 2018; Zhou *et al.*, 2020] where graphs serve as spatial domain discretization.

Several vertex ordering techniques have been proposed in the literature (refer to Section 2), leveraging both topological properties and spatial proximity to produce optimal vertex ordering. Many of these methods focus on minimizing index discontinuities by reordering nodes according to edge connectivity or employing dimensionality reduction techniques to project spatial relationships into a linear ordering. This applies to graphs derived from geolocated entities such as street networks [Salinas *et al.*, 2022], census tract graphs [Garcia-Zanabria *et al.*, 2019], and mobility graphs [Pan *et al.*, 2022], which are commonly used in urban data analysis. However, evaluating the effectiveness of these approaches remains challenging, as traditional metrics often rely on global quality assessments that reduce the order quality to a single numerical value.

Although some studies have compared vertex ordering methods using different metrics [Behrisch *et al.*, 2016], few have specifically addressed the challenge of comparing topology and geometry-based approaches for spatial graphs. A major limitation in such comparisons is the lack of appropriate evaluation methodologies that integrate both qualitative and quantitative analysis, particularly at a local spatial level. Existing assessment methods often fail to fully capture how different ordering techniques perform, making it difficult to determine their suitability for specific applications.

In previous work [Salinas *et al.*, 2024], we proposed a visual evaluation methodology that combines two quantitative geometric approaches, called Forward and Inverse, with qualitative methods for assessing vertex ordering. These qualitative methods include histograms, boxplots, and spatial maps that illustrate ordering patterns, relying on a color scale to depict the spatial distribution of ordered indices, while highlighting discontinuities in the ordering process. Quantitative metrics have also been employed, enabling valuable evaluation focused on spatial relationships without explicitly considering topological properties.

This work builds upon the previous methodology proposed by Salinas et al. [Salinas *et al.*, 2024], introducing two novel topological approaches for visually evaluating vertex ordering algorithms in spatial graphs. By incorporating topological measures, we complement and compare the previously proposed Forward and Inverse evaluation methods, offering a more comprehensive framework to assess how well an ordering preserves spatial and topological structures. Furthermore, our approach streamlines the evaluation process, significantly improving computational efficiency. We validate our methodology through experiments on urban datasets from multiple cities, demonstrating its effectiveness in capturing spatial and structural distortions that global metrics may overlook.

**Contributions.** The main contributions of this work are:
- The introduction of two novel topological approaches for evaluating the quality of graph vertex ordering.
- A comparative analysis of Geometric Forward, Geometric Inverse, Topological Forward and Topological Inverse Approaches to vertex ordering.
- A comprehensive categorization of related work on vertex ordering, distinguishing between ordering techniques and evaluation metrics.
- A comprehensive set of experiments demonstrating how the proposed visual methodology enhances the assessment of vertex ordering methods in urban applications.

## 2 Related Work

This section provides an overview of vertex ordering methods and metrics commonly used to evaluate such techniques. Specifically, we focus on methods that strive to minimize index jumps between neighboring vertices. A more comprehensive discussion about vertex ordering methods and their evaluation can be found in [Behrisch *et al.*, 2016] and [Mueller *et al.*, 2007] surveys.

### 2.1 Vertex ordering methods

Vertex ordering methods have been studied since the 1950s [Robinson, 1951], originally in the context of matrix reordering [Rosen, 1968; Cuthill and McKee, 1969]. To validate and demonstrate the potential of the proposed visual methodology, we assessed three vertex ordering techniques: Fiedler vector-based [Salinas *et al.*, 2024], t-SNE [Van der Maaten and Hinton, 2008], and UMAP [McInnes *et al.*, 2018]. All these alternatives, which are briefly described in the following, compute a 1D embedding of the vertices, which is then exploited to infer their ordering.

#### 2.1.1 Geometric Approaches (Based on Node Spatial Location)

For geolocated graphs, space-filling curves (SFCs) have been widely used to traverse the spatial domain, mapping nodes onto a one-dimensional (1D) sequence based on the curve's traversal [Mark, 1990; Morton, 1966; Mokbel *et al.*, 2003]. Another common strategy is dimensionality reduction, which projects spatial graphs into a lower-dimensional space to infer an ordering. Methods such as Multidimensional Scaling (MDS) [Hamer and Young, 2013], t-SNE [Van der Maaten and Hinton, 2008], and UMAP [McInnes *et al.*, 2018] have been employed for this purpose, as well as more general dimensionality reduction techniques [Nonato and Aupetit, 2018].

A related approach involves graph embedding methods, where graphs are first embedded into a high-dimensional space and then projected onto 1D using Principal Component Analysis (PCA) [Harel and Koren, 2002; Elmqvist *et al.*, 2008]. Graph Neural Networks (GNNs) [Zhang *et al.*, 2019] have also been employed, learning node embeddings that can be used for ordering. Matrix completion techniques [da Fontoura Costa, 2023] estimate missing structural information to optimize the ordering process.

**t-SNE** is widely used for visualizing high-dimensional data in lower-dimensional spaces. It measures the similarity between data points by their likelihood of being neighbors, considering their pairwise distances in the high-dimensional space [Van der Maaten and Hinton, 2008]. t-SNE is a powerful dimensionality reduction technique, it constructs a layout where nearby points in the original space are mapped closer to each other in the lower-dimensional space. The number of neighbors is a hyperparameter that substantially influences the outcomes of the algorithm, namely the perplexity. Formally, the t-SNE algorithm starts by computing pairwise similarities between data points in the high-dimensional space using a Gaussian kernel as a probability density function:

$$p_{j|i} = \frac{\exp(-||x_i - x_j||^2/(2\sigma_i^2))}{\sum_{k \neq i} \exp(-||x_i - x_k||^2/(2\sigma_i^2))},$$

where $x_i$ and $x_j$ are the spatial coordinates of the graph nodes. Points are then placed in the low-dimensional space by minimizing the Kullback-Leibler (KL) divergence between $p_{j|i}$ and a Student's t-distribution $q_{j|i}$ in the lower-dimensional space. For vertex ordering, this space is set to 1D.

**UMAP** is a dimensionality reduction technique based on topological data analysis and manifold learning. It provides a good low-dimensional representation by considering a graph that captures the topology of the source data. Similar to t-SNE, UMAP is also a $k$-neighbor graph learning algorithm. However, while t-SNE tends to preserve the local structure of the data, UMAP strives to preserve the local and global structure. The algorithm starts by constructing a particular weighted $k$-neighbor graph. For each input point $x_i$, it obtains a local graph $\overline{G} = (V, E, w)$, where $V$ are the set of vertices, $E$ directed edges and $w$ is an edge weight function:

$$w((x_i, x_j)) = \exp\left(\frac{-\max(0, d(x_i, x_{i_j}) - \rho_i)}{\sigma_i}\right),$$

where $d(\cdot, \cdot)$ is the distance in the original space, $\rho_i$ is the minimal distance between the node $i$ and its neighbors, and $\sigma_i$ is a normalization factor (see [McInnes *et al.*, 2018] for details). The edge weight function derives from finding the $k$ nearest neighbors of each $x_i$ [Dong *et al.*, 2011], and such local graphs are combined into a unified topological representation using the probabilistic t-conorm. UMAP constructs a low-dimensional layout of the entire graph using a force-directed graph layout mechanism. In our case, the mapping is performed into a 1D space, from which the vertex order is derived.

In this work, t-SNE and UMAP are used to map the coordinates of the vertices of an undirected connected spatial (UCS) graph to a 1D space, that is, mapping is performed from $S$ (a 2D space) to $\mathbb{R}$.

### 2.1.2 Topological Approaches

A different class of methods relies on the topological properties of the graph rather than its spatial attributes. Spectral methods, particularly those using the Fiedler vector, have proven effective in ordering graph vertices based on connectivity patterns [Atkins *et al.*, 1998]. These methods have been applied in various domains, such as seriation [Concas *et al.*, 2023] and ranking problems [Chau *et al.*, 2022]. Another topological approach is Cauchy graph embedding, which aims to preserve local graph topology [Luo *et al.*, 2011]. Although built upon a solid mathematical framework, Cauchy graph embedding is computationally intensive, hampering its use for large-scale graphs. Heuristic strategies have also been explored, such as iterating through adjacency matrix rows/columns to optimize index assignments [Niermann, 2005] or transforming adjacency matrices into simpler representations [Mafteiu-Scai, 2014]. Additionally, graph theory applied to other fields, such as chemistry to model molecular structures [Camacho *et al.*, 2020] and in clustering problems, has also been adapted to vertex ordering [Costa and Tokuda, 2022]. These applications further demonstrate the versatility of topological techniques in structuring and analyzing complex datasets.

**Fiedler:** The Fiedler vector can be utilized as a basis for vertex ordering where the nodes are ordered according to the Fiedler vector values. Specifically, in complex networks visualization, the Fiedler vector has been employed to sort the nodes of a network when building matrix-based representations. Let $\mathcal{G}$ be a connected graph, and let $\mathbf{L}$ denote its Laplacian matrix. In our context, $\mathcal{G}$ represents the street map graph of a city, where the street corners correspond to the vertices and street blocks connecting two corners correspond to the edges of $\mathcal{G}$. Let $i$ and $j$ be two distinct vertices of $\mathcal{G}$. The Laplacian matrix $\mathbf{L}$ is built by setting $L_{ij} = -1/l_{ij}$ if $i$ and $j$ are adjacent, where $l_{ij}$ is the length of the corresponding edge (length of the street block connecting); $L_{ij} = 0$ if $i$ and $j$ are not adjacent; and $L_{ii} = \sum_{j \neq i} |L_{ij}|$ ($|\cdot|$ is the absolute value). Since the Laplacian matrix $\mathbf{L}$ is symmetric and positive semi-definite, their eigenvalues are real and non-negative. Moreover, when $\mathcal{G}$ is connected, the Laplacian matrix $\mathbf{L}$ has exactly one eigenvalue equal to zero. The Fiedler vector is the eigenvector of $\mathbf{L}$ associated with the smallest non-zero eigenvalue. For further details, consult Chung's book [Chung and Graham, 1997]. Let $\mathbf{v} = [v_1, \ldots, v_n]$ denote the Fiedler vector of $\mathcal{G}$, where $n$ is the number of nodes in $\mathcal{G}$. The entry $v_i$ of $\mathbf{v}$ corresponds to the node $i$ in $\mathcal{G}$, meaning that $\mathbf{v}$ maps each node to a real number. The Fiedler ordering rearranges the nodes of $\mathcal{G}$ based on their corresponding values $v_i$. This ordering (and the Fiedler vector) is determined by minimizing the expression:

$$\mathbf{v} = \min_{\|u\|=1} \sum_{\substack{\forall\, i \neq j \text{ s.t.} \\ L_{ij} \neq 0}} |L_{ij}|(v_i - v_j)^2. \qquad (1)$$

Equation (1) reveals that pairs of adjacent nodes $i$ and $j$ tend to hold similar values $v_i$ and $v_j$. Also, the similarity depends on the spatial proximity of nodes due to the weights $|L_{ij}|$. Thus, the resulting embedding is expected to preserve the relative distances between adjacent (and neighboring) nodes.

## 3 Comparing Vertex Ordering Algorithms

In this work, we advance the visual methodology for evaluating vertex ordering methods in undirected connected spatial (UCS) graphs proposed in [Salinas *et al.*, 2024]. Specifically, we propose two novel topological approaches to locally measure the quality of vertex ordering algorithms.

Consider a UCS graph $\mathcal{G} = ((V, s), E)$, where $V$ is the set of $n$ vertices, $s \colon V \to S$ maps each vertex to its spatial coordinates, and $E$ represents the set of edges. The space $S$ corresponds to the geographic coordinate system of the vertices in $V$, which in our case is $\mathbb{R}^2$. An ordering is formally defined as a bijection $\phi(v) \to \{1, \ldots, n\}$, where each vertex $v \in V$ is assigned a unique index. Our methodology explicitly considers the spatial coordinates of UCS graphs to assess vertex ordering techniques, which can enhance various spatial data analysis tasks.

In previous work [Salinas *et al.*, 2024], we introduced two local quantitative measures: the Geometric Forward Approach, which assumes that nodes with similar indices in the 1D ordering should also be spatially close; and the Geometric Inverse Approach, which considers that nodes within a given spatial region should also be close in the 1D ordering. Both approaches are defined for each vertex of the graph.

While these approaches effectively capture spatial coherence, they do not explicitly account for the structural properties of the graph. To address this gap, we now extend our analysis by introducing two topological approaches, which evaluate how well the ordering preserves topological structures. These novel approaches allow for a direct comparison with the previously proposed Forward and Inverse methods. Specifically, it considers:

- Global connectivity patterns, ensuring that the ordering maintains long-range dependencies within the graph.
- Structural consistency, preserving clusters and hierarchical relationships in the 1D ordering.
- Optimization of spatial and structural coherence, balancing local spatial proximity with topological fidelity.

These approaches enable the comparison of different node embedding techniques. Notably, graph-based visual analytic tools can greatly benefit from these orderings: by structuring the vertices in a meaningful sequence, visual representations can more intuitively reveal complex relationships and patterns. Proper vertex ordering facilitates the identification of clusters, hierarchies, and trends, allowing users to derive insights more efficiently and make informed decisions.

## 3.1 Geometric Forward Approach

The Geometric Forward Approach, originally named Forward Approach in [Salinas *et al.*, 2024], evaluates the quality of a vertex ordering based on local compactness in the spatial domain. To this end, a window of size $m$ is defined around each vertex in the given ordering. Compactness is measured using the diagonal length of the bounding box $B_{i_k}$ in the original space containing the vertices within the window. This length is then normalized with respect to an optimal bounding box $\tilde{B}_{i_k}$ constructed from the $m$ nearest neighbors of the vertex in the spatial domain. Finally, a sliding window approach is applied throughout the ordering sequence, generating a set of diagonal length values, where smaller values indicate a better ordering.

In Figure 1(a), the Geometric Forward Approach is applied to the street map of São Paulo, where nodes (representing street corners) are ordered using the Fiedler method. The evaluation of these nodes is performed locally using the sliding windows, as illustrated at the bottom in Figure 1(a). The green window corresponds to a set of vertices, represented on the map, resulting in a small bounding box with a diagonal length of 3.34 km. On the other hand, the red window leads to a much larger and undesirable bounding box with a diagonal length of 28.93 km. This figure is adapted from [Salinas *et al.*, 2024].

## 3.2 Geometric Inverse Approach

The Geometric Inverse Approach, originally called Inverse Approach in [Salinas *et al.*, 2024], assesses the quality of a vertex ordering based on how well the spatial locality of nodes is preserved in the ordering sequence. Given a vertex $v_i$, its neighborhood $R_i(r)$ is defined as the set of vertices within a graph distance $r$. The corresponding indices of these vertices in the ordering are collected in $I_{R_i(r)}$. The ordering quality is then quantified by the normalized diameter of this index set, given by

$$d_i^{inv} = \frac{\max(I_{R_i(r)}) - \min(I_{R_i(r)})}{|R_i(r)|},$$

where $\max(I_{R_i(r)})$ and $\min(I_{R_i(r)})$ denote the largest and smallest indices in $I_{R_i(r)}$, respectively, and $|R_i(r)|$ is the number of nodes in the neighborhood. A global assessment is obtained by computing these values for all nodes, forming a set $D^{inv}$, where smaller values indicate a better ordering.

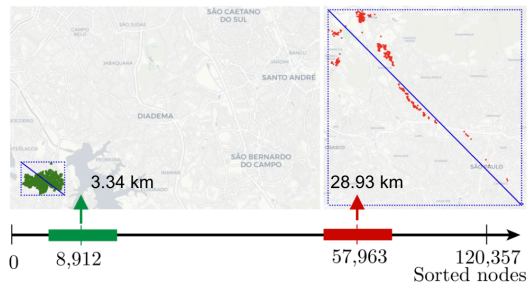

(a) Geometric Forward Approach

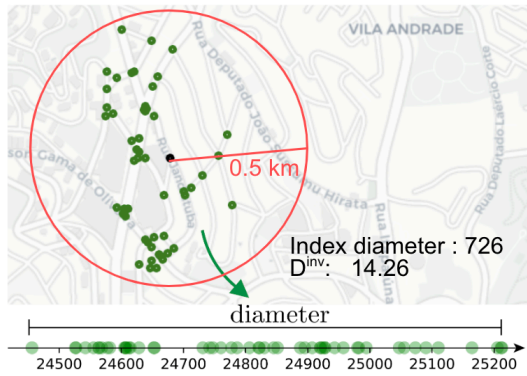

(b) Geometric Inverse Approach

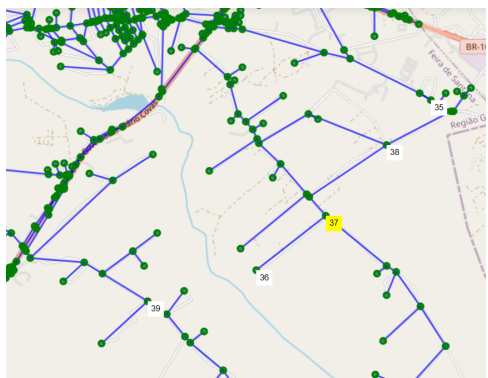

(c) Topological Forward Approach

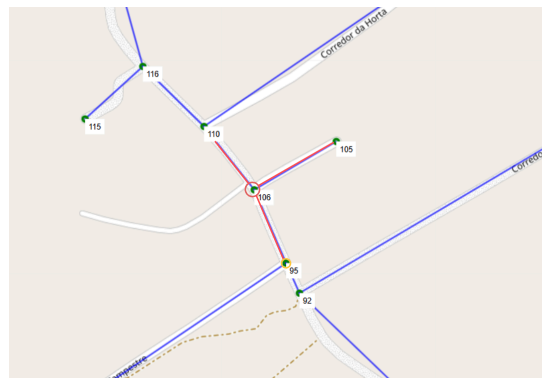

(d) Topological Inverse Approach

**Figure 1.** Illustration of the four metric approaches applied to street maps: (a) Geometric Forward, which evaluates local compactness using sliding windows; (b) Geometric Inverse, which assesses the ordering quality based on the distance to neighboring nodes within a fixed radius; (c) Topological Forward, which evaluates the number of hops between neighbors in the ordering; and (d) Topological Inverse, which measures the spread of indices among a node's neighbors.

As shown in Figure 1(b), the Geometric Inverse Approach uses a reference node (denoted by the black dot) and considers all nodes within a radius of $r = 0.5$ km, as indicated by the red circle. The selected neighboring nodes are highlighted in green. The quality of the vertex ordering is evaluated by measuring the diameter of the range of the corresponding index set $I_{R_i(r)}$, which in this case has a diameter of 726. This value is then normalized by the number of nodes within that range, resulting in the normalized measure $d_i^{inv} = 14.26$. Although this metric is shown for the black reference node, it can be iteratively computed for all graph nodes $v_i$, generating a set of diameters, denoted as $D^{inv}$, for a fixed radius $r$. This figure is adapted from [Salinas *et al.*, 2024].

### 3.3 Topological Forward Approach

The Topological Forward Approach measures whether vertices with similar order are close to each other in the graph. Similar to the Geometric Forward Approach (Section 3.1), a window of size $m_i$ is defined around each vertex $v_i$ in the given ordering, giving rise to a set of vertices $I_{m_i}(v_i)$. Instead of relying on geometric information to estimate the compactness of $I_{m_i}(v_i)$, we focus on the graph distance, *i.e.*, the shortest path between vertices. Accordingly, for each vertex $v_i$ in a UCS graph $\mathcal{G}$, we define the forward topological measure as:

$$t_i = \max_{v_j \in I_{m_i}(v_i)} d(v_i, v_j),$$

where $d(v_i, v_j)$ denotes the shortest path distance between $v_i$ and $v_j$.

The window size $m_i$ is a key parameter that directly influences the measure $t_i$ and adapts to each vertex according to its degree. In particular, vertices with higher degrees are expected to be evaluated over a larger window in the ordering. We thus define $m_i$ as:

$$m_i = \begin{cases} |N(v_i)|, & \text{if } |N(v_i)| \text{ is even,} \\ |N(v_i)| + 1, & \text{if } |N(v_i)| \text{ is odd.} \end{cases}$$

As illustrated in Figure 1(c), consider a node that is assigned index 37 according to a given ordering method. This node has three immediate neighbors in the graph. To compute its Topological Forward value, we measure the distance to other nodes in terms of hops, where a hop corresponds to traversing a single edge along the shortest path connecting two nodes. For instance, a node directly connected to the reference node is at a distance of one hop; a node reached by passing through one intermediate node is at a distance of two hops, and so on. In this example, we take two nodes before it in the ordering (35 and 36) and two after (38 and 39). We then measure the number of hops required to reach each of these nodes from node 37. The hop counts are: 3 (to 35), 1 (to 36), 2 (to 38), and 29 (to 39). The highest of these values — 29 in this case — is the Topological Forward value for node 37.

### 3.4 Topological Inverse Approach

The Topological Inverse Approach measures whether the index of a vertex, in the given ordering, is close to that of its neighbors in the graph. In other words, it measures how much a vertex's index deviates from the index of its immediate neighbors.

Let $\mathcal{G}$ be a UCS graph and let $\phi$ be the ordering of the vertices, obtained from a vertex ordering algorithm. The difference between the indices of two neighboring vertices $v_i$ and $v_j$ is given by $|\phi(v_i) - \phi(v_j)|$. Since each vertex $v_i$ may have multiple neighbors, we quantify local consistency by considering the maximum difference among all neighbors. More precisely, we define the inverse topological measure as

$$t_i^{inv} = \max_{v_j \in N(v_i)} \frac{|\phi(v_i) - \phi(v_j)|}{|N(v_i)|},$$

where $N(v_i)$ is the set of all vertices directly connected to $v_i$ by an edge. Thus, each value $t_i^{inv}$ is a real number that effectively quantifies local inconsistencies in the vertex ordering around vertex $v_i$, based on the topology of the graph. A lower $t_i^{inv}$ value indicates that the ordering preserves local consistency, meaning a vertex's index is similar to those of its neighbors. In contrast, a higher $t_i^{inv}$ suggests greater deviation between a vertex's index and those of its neighboring vertices.

As shown in Figure 1(d), each node is represented by a green point, each edge by a blue line, and the indices assigned by the ordering technique are marked in the white box. Our reference node, highlighted with a red circle, has index 106, and its immediate neighbors, connected by red lines, are labeled with indices 110, 105, and 95. Among these neighbors, the most distant is node 95, emphasized in yellow. The largest distance between the reference node and the most distant neighbor is calculated as $|106 - 95| = 11$. Since the reference node has three immediate neighbors, the value is normalized, and the resulting topological measure for this node is $11/3 = 3.6666$. This measure quantifies the spread of indices among the neighboring nodes, providing an assessment of the vertex ordering.

## 4 Experimental Results and Interpretation

We conducted experiments to evaluate the effectiveness of the proposed methodology using three vertex ordering techniques: Fiedler, t-SNE, and UMAP, briefly described in Section 2.1. Our evaluation framework combines the Geometric and Topological (Forward and Inverse) Approaches outlined in Section 3, supplemented by visual qualitative analysis.

To ensure a diverse and representative experimental setup, we selected street map graphs from cities with varying characteristics. Specifically, we used street graphs from São Paulo, Maceió, Barcelona, Busan, Mumbai, Nairobi, and Bogotá, obtained from OpenStreetMap (OSM, [OpenStreetMap contributors, 2017]). In the Geometric Forward Approach, we used a window size of $n/100$, while the Geometric Inverse Approach employed a radius of $0.5 km$.

For benchmarking, we compared the performance of the vertex ordering techniques against two baselines: the original OSM vertex order and a random vertex order.

First, we demonstrate how our methodology can be applied to select suitable ordering techniques for specific cities.

Next, we show how to use our approach to visually assess regions with coherent and inconsistent ordering indices. Finally, we illustrate how our measures capture performance variations depending on the hyperparameter values used by the ordering techniques.

## 4.1 Using the measures to compare ordering techniques

Figure 2 compares the performance of various ordering techniques across multiple cities (Barcelona, Busan, Mumbai, Nairobi, and Bogotá) using a single run without hyperparameter tuning. Each boxplot represents the local performance for each node.

In general, t-SNE performs better on the Geometric Forward Approach, except in Barcelona, where UMAP outperforms it. For the Geometric Inverse Approach, t-SNE also performs better, though the advantage is less pronounced. On the Topological Inverse measure, Fiedler, t-SNE, and UMAP show similar median values in most cities (except Barcelona). However, Fiedler often reaches slightly lower values, suggesting that it provides a more suitable ordering for certain nodes—likely due to its reliance on graph topology.

The original and random orderings show the worst performance across most metrics and cities. An exception is the topological forward measure, where the original ordering has a consistently low median, despite a wide value spreading. A likely explanation for this behavior is rooted in the order in which nodes are included: the original OSM ordering often reflects the sequence in which nodes were added to the graph. We believe nodes are included in chunks of compact areas, resulting in consecutive indices. This pattern creates small, densely ordered, and sometimes disconnected clusters that lead to lower Topological Forward scores. Nonetheless, despite the apparent advantage in this single metric, the overall performance of the original ordering remains poor across other measures, underscoring that excellence in one measure does not necessarily ensure comprehensive performance.

These results demonstrate how our methodology can be directly applied to select an appropriate vertex ordering technique for a specific city from multiple perspectives. Next, we explore how these measures relate to the vertex ordering.

Figure 3 presents the quantitative and qualitative results for different vertex ordering techniques applied to the Maceió street graph. The first column shows the quantitative performance for each technique using the four measure approaches (geometric forward, geometric inverse, topological forward, and topological inverse). The second column provides qualitative insights by displaying the vertex orderings estimated by each technique, with nodes color-coded according to their indices using the shown color scale.

Consistent with the results from other cities, the original and random orderings perform the worst across most measures, except for the topological forward measure. t-SNE performs slightly better on the Geometric Forward and Inverse Approaches, while Fiedler outperforms the others on the Topological Inverse Approach.

The color-coded visualizations reveal that each ordering technique produces distinct patterns. Fiedler's ordering shows smooth transitions from southeast to northwest. However, vertices on the far east and far west share similar colors, indicating that spatially distant vertices receive close indices. This explains why Fiedler did not outperform t-SNE and UMAP on the forward measure. Additionally, Fiedler consistently shows a smaller range of ordering indices compared to other techniques. In other words, the distribution of indices produced by Fiedler is more compressed, with fewer extreme values, due to its smooth transitions that minimize abrupt changes and outliers.

In contrast, t-SNE produces a more clustered pattern with noticeable color (and index) discontinuities, particularly between yellow and blue regions. These discontinuities occur between clusters, but within each cluster, the ordering is more consistent than with Fiedler, leading to better overall performance according to the median.

Similar to Fiedler, UMAP exhibits smooth transitions in the denser central areas of the map but introduces discontinuities in less dense regions, similar to t-SNE.

The original OSM ordering reflects the order in which nodes were added to the graph, from dark blue (older nodes) to red (newer ones). This results in central areas dominated by dark blue tones, with nearby shades grouped together.

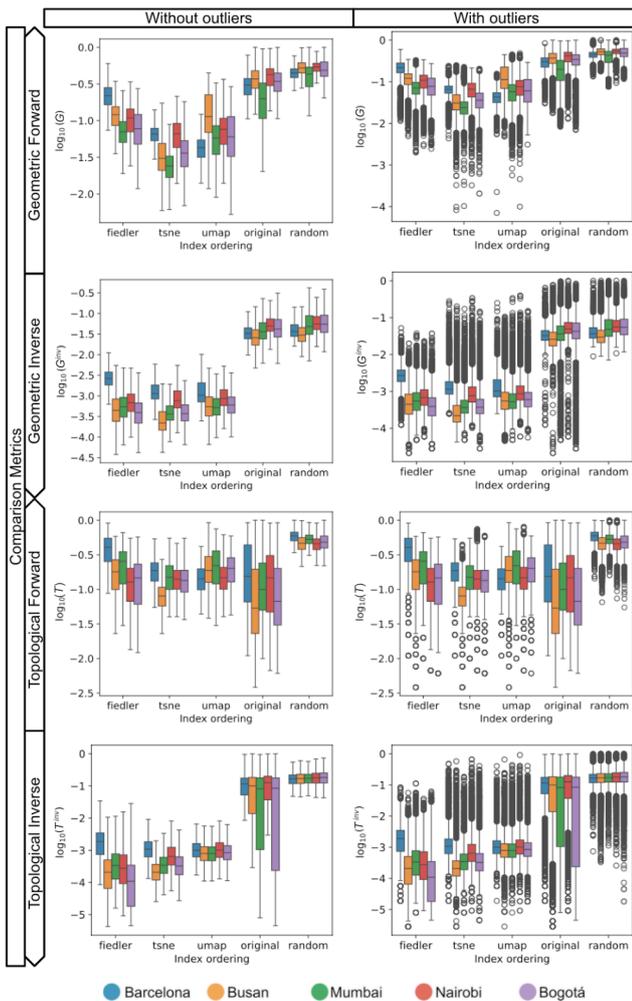

**Figure 2.** General comparison of techniques in several cities. From top to bottom, the figure shows the results for the approaches: geometric forward, geometric inverse, topological forward, and topological inverse. The first column of the figure displays the performances without outliers, while the second column includes them. All boxplots are presented on a logarithmic scale and are normalized per city for better visualization.

Across the city, we observe small, disconnected clusters in similar colors, indicating compact but isolated groups of consecutively indexed nodes.

## 4.2 Using Measures to Identify Errors and Inconsistencies

Beyond analyzing the performance of each ordering method, our methodology enables the identification of regions where the ordering is more or less consistent. To illustrate this application, we examine the measures and techniques using the São Paulo street graph. The left column in Figure 4 shows the performance of the vertex ordering techniques for the four measures—Geometric Forward, Geometric Inverse, Topological Forward, and Topological Inverse.

t-SNE performs better overall, outperforming other techniques on all measures except Topological Forward. Fiedler slightly outperforms UMAP on the geometric measures, and this is most evident on the Topological Inverse Approach, likely due to Fiedler's use of the street graph's topology in its ordering process. In contrast, the original and random orderings perform the worst across most measures, with the exception of the original ordering on the Topological Forward measure, for reasons discussed earlier.

Since t-SNE showed a good overall performance, we further analyzed its ordering index by comparing it to the errors indicated by the measures.

In Figure 4, the first column shows boxplots for each ordering method, where each row corresponds to one of the four metrics. The second column presents color-coded maps showing the geographic distribution of error values for the t-SNE ordering, allowing spatial interpretation of its performance. Here, the error values correspond to the metric in the respective row (e.g., Geometric Forward in the first row, Geometric Inverse in the second, and so on), and are computed by comparing the t-SNE ordering against the ground-truth graph structure. Additionally, Figure 5 shows a color-coded map of the t-SNE ordering itself, enabling a direct visual comparison between the order and the corresponding error patterns across the city.

Both inverse approaches highlight the discontinuity borders produced by t-SNE, but in different ways. The Geometric Inverse Approach marks these borders with thicker lines due to its radius-based measurement, while the Topological Inverse Approach pinpoints more local discontinuities related to immediate graph neighbors. Moreover, the Geometric Forward Approach highlights regions where neighboring nodes in the vertex ordering are geographically distant, revealing areas of poor local ordering. Similarly, the Topological Forward measure highlights a poorly ordered region in the south of the city (bright red). As shown in the zoomed view in Figure 6, this area is split by a river. Although t-SNE assigns similar indices due to geographic proximity, the street network reveals that the two sides are mostly disconnected due to the lack of bridges or direct connections over the river.

Our evaluation method provides a comprehensive perspective on vertex ordering reliability, revealing complementary insights into the local behavior of the ordering techniques.

## 4.3 Hyperparameter analysis

Another potential application of our methodology is optimizing hyperparameters for vertex ordering techniques. To demonstrate this, we analyzed the impact of the perplexity hyperparameter in t-SNE using the São Paulo street graph. We tested eight perplexity values: 5, 10, 25, 50, 75, 100, 150, and 200. The right column in Figure 7 shows the qualitative results for four perplexity values, with city nodes (corners) color-coded according to their indices.

The visualization reveals distinct patterns for different perplexity values. For example, a perplexity of 10 shows more discontinuities, with darker blue shades in the extreme northwest and orange in the extreme north also appearing in the bottom of the northeast region. With perplexity 50, the northwest and extreme north regions become lighter. At 100, transitions in the central part of the map are a bit smoother, and at 200, nodes with similar orange tones are placed closer together.

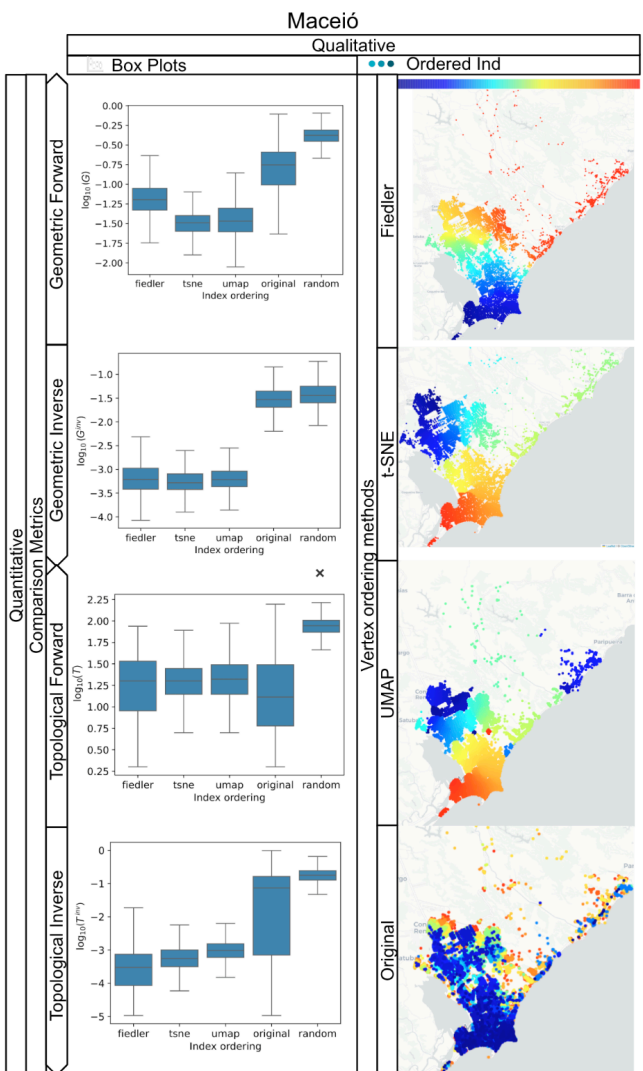

**Figure 3.** Performance analysis of all ordering methods in Maceió using boxplots and detailed visualizations of their ordering indices. In the first column, each row corresponds to one metric (from top to bottom: Geometric Forward, Geometric Inverse, Topological Forward, and Topological Inverse). The second column shows a map for each ordering method (from top to bottom: Fiedler, t-SNE, UMAP, and the original OSM order), where nodes are color-coded by their assigned order—from low (dark blue) to high (red), passing through intermediate values (light blue, green, yellow).

Boxplots on the left in Fig. 7 indicate that t-SNE's performance improves with increasing perplexity, as evidenced by shorter boxplots in the geometric forward measure, following an exponential trend. Similar reductions in dispersion are, in fact, seen in all measures.

It is important to note that although the performance metrics improve, the results are still not perfect, as the measures indicate remaining errors. This suggests room for further refinement, which could be achieved by adjusting other parameters or exploring alternative vertex ordering techniques.

This experiment illustrates how the proposed approach can assist in fine-tuning hyperparameters for vertex ordering methods, as demonstrated in the case of selecting a perplexity value for t-SNE. By providing four distinct perspectives on what defines a good ordering, the analysis highlights patterns that would be difficult to detect by visual inspection alone.

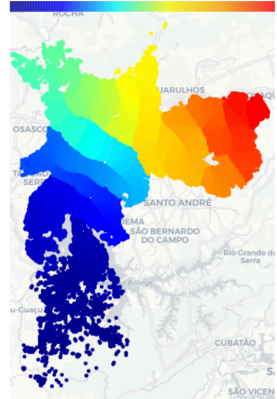

**Figure 5.** São Paulo map showing the t-SNE ordering, with node indices color-coded from low (dark blue), through mid-range (light blue, green, yellow), to high (red).

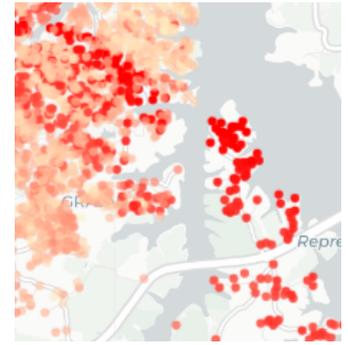

**Figure 6.** South of São Paulo zoomed in to show local low performance according to the Topological Forward Measure.

## 5 Practical Application of Vertex Ordering

While our work focuses on the systematic evaluation of vertex ordering algorithms, it is useful to highlight a concrete application where such ordering has a practical impact. A notable example is the study by Santos *et al*. [2024], which applies 1D vertex ordering to space-time crime data in Maceió, Brazil. Their work demonstrates how vertex ordering can reveal recurring patterns and trends that would be difficult to detect otherwise, providing strong evidence of the method's practical utility. Figure 8 (adapted from Santos *et al*. [2024]) shows a space-time heatmap of quarterly aggregated crimes in Maceió, Brazil, from 2019 to 2021 (left). Each column corresponds to a specific location in the city (graph vertices), depicting crime rates over time. The locations are sorted using 1D t-SNE and are arranged accordingly in the heatmap, thus making nearby locations represented as neighboring columns in the heatmap. The rows account for the quarterly aggregated intensity of crimes in each location. Two regions, delineated by dashed green rectangles, deserve special attention due to their consistent crime behavior over time, demonstrating how vertex ordering facilitates the identification of recurrent spatial-temporal patterns in crime data. The corresponding locations are highlighted in the map on the right.

## 6 Conclusion

This paper presented a methodology for spatial graph vertex ordering, introducing two new topology-based metrics and a comparative framework. By combining geometric and topological measures, our approach provides a more detailed understanding of ordering performance across different urban networks.

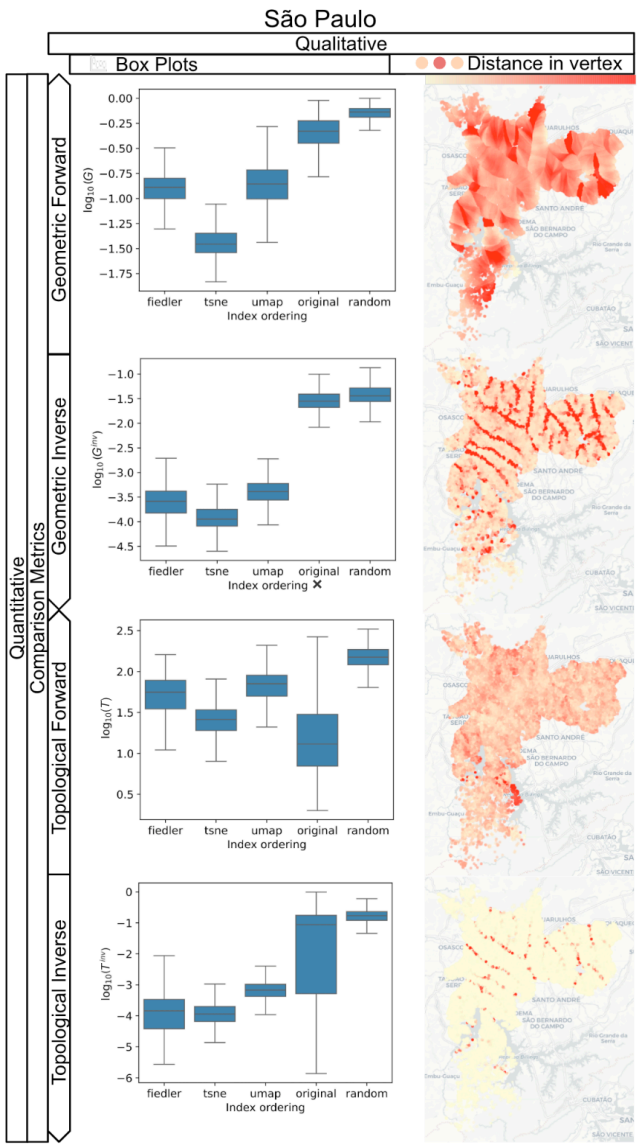

**Figure 4.** Performance analysis of all ordering methods in São Paulo using boxplots, with a detailed view of t-SNE performance via maps. Each row corresponds to one metric (from top to bottom: Geometric Forward, Geometric Inverse, Topological Forward, and Topological Inverse). The first column shows boxplots for all methods. The second column displays maps where each node is color-coded by its t-SNE error value for the corresponding metric, ranging from beige (low values, better performance) to red (high values, worse performance).

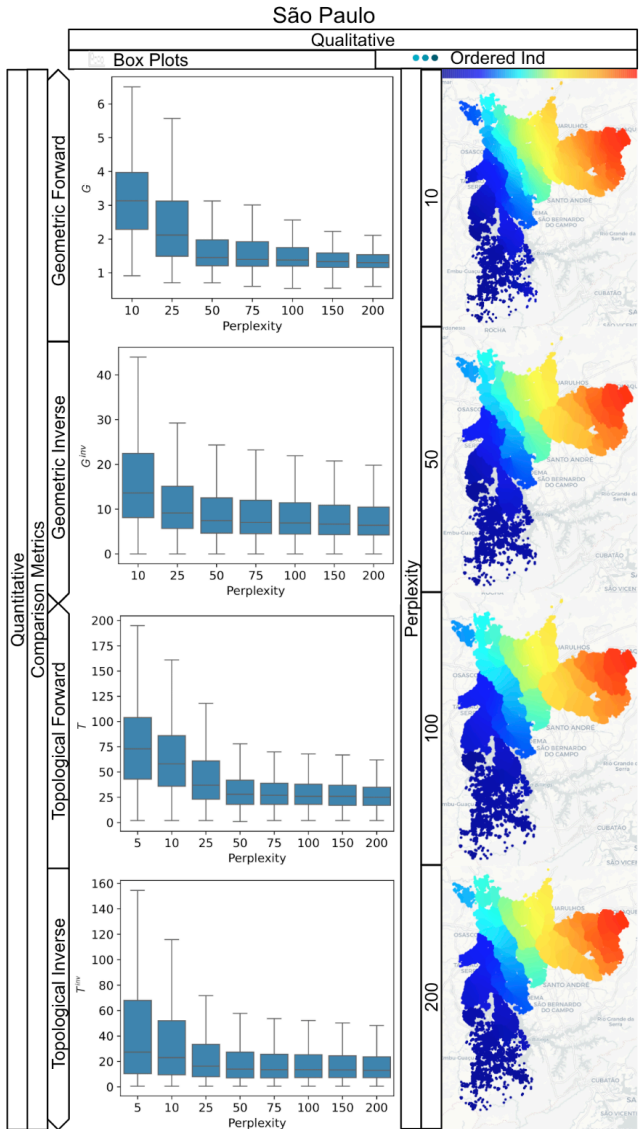

**Figure 7.** Example analysis of t-SNE performance across different perplexity values for ordering the nodes of the São Paulo street graph. The first column shows boxplots for each perplexity setting across the four measures (from top to bottom: Geometric Forward, Geometric Inverse, Topological Forward, Topological Inverse). The second column presents color-coded maps of the t-SNE orderings, each using a different perplexity value (from top to bottom: 10, 50, 100, 200).

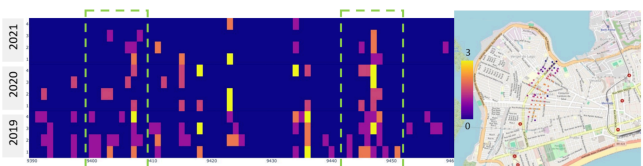

**Figure 8.** Space-time heatmap displaying quarterly aggregated crimes in Maceió, Brazil, from 2019 to 2021. The horizontal axis represents graph vertices (city areas), and the vertical axis represents time in quarterly intervals. Two regions, delineated by green rectangles, show consistent crime behavior over time. Figure adapted from Santos *et al.* [2024].

Our experiments showed how the methodology can be used to select suitable vertex ordering techniques, fine-tune hyperparameters, and identify regions with varying ordering reliability. The different measures highlighted complementary aspects of performance, giving a more complete picture of each technique's behavior.

Future work could explore applying this methodology to other types of graphs and experimenting with additional metrics that account for dynamic graph properties. Investigating the effects of different spatial resolutions and graph granularities may also provide further insights.

Overall, this approach offers a useful tool for evaluating vertex ordering methods, supporting more informed choices in graph analysis tasks.

## Declarations

### Authors' Contributions

Karelia Salinas: Writing, Visualization; Victor Barella: Investigation (experiments); Thales Vieira: Formal Analysis (formalization of the Topological Approach); Gustavo Nonato: Methodology (formulation of the new approach), Supervision, Writing – review and editing. All authors read and approved the final manuscript.

### Competing interests

The authors declare that they have no competing interests.

### Acknowledgements

This work was supported by FAPESP (#2020/07012-8, #2022/09091-8, #2023/15805-6) and CNPq (#307184/2021-8). The opinions, hypotheses, conclusions and recommendations expressed in this material are the responsibility of the authors and do not necessarily reflect the views of FAPESP and CNPq.

### Availability of data and materials

The datasets (and/or softwares) generated and/or analysed during the current study are available in https://github.com/giva-lab/vertex_ordering.